\documentclass[]{aa}

\usepackage{graphicx}
\usepackage{txfonts}
\usepackage{psfig}

\begin{document}

\title{On the interaction of microquasar jets with
stellar winds}
\titlerunning{Interaction of $\mu Q$ jets with stellar winds}
\author{M. Perucho\inst{1} \and V. Bosch-Ramon\inst{2}}
\institute{Max-Planck-Institut f\"ur Radioastronomie, Auf dem
H\"ugel, 69, Bonn 53121, Germany; perucho@mpifr-bonn.mpg.de \and
Max Planck Institut f\"ur Kernphysik, Saupfercheckweg 1,
Heidelberg 69117, Germany; vbosch@mpi-hd.mpg.de}

\offprints{ \\ \email{perucho@mpifr-bonn.mpg.de}}

\abstract{Strong interactions between jets and stellar winds at
binary system spatial scales could occur in high-mass microquasars.}{We
study here, mainly from a dynamical but also a radiative
point of view, the collision between a dense stellar wind and
a mildly relativistic hydrodynamical jet of supersonic nature.}{We
have performed numerical 2-dimensional simulations of jets, with
cylindrical and planar (slab) symmetry, crossing the stellar wind
material. From the results of the simulations, we derive
estimates of the particle acceleration efficiency, using first
order Fermi acceleration theory, and give some insight on the
possible radiative outcomes.}{We find that, during jet launching,
the jet head generates a strong shock in the wind. During and
after this process, strong recollimation shocks can occur due to
the initial overpressure of the jet with its environment. The
conditions in all these shocks are convenient to accelerate
particles up to $\sim$~TeV energies, which can lead to leptonic
(synchrotron and inverse Compton) and hadronic (proton-proton)
radiation.
In principle, the cylindrical jet simulations show that
the jet is stable, and can escape from the system even for
relatively low power. However, when accounting for the wind
ram pressure, the jet can be bent and disrupted for power
$\la 10^{36}$~erg~s$^{-1}$.}{}


\maketitle

\section{Introduction} \label{intro}

Strong shocks take place in very different scenarios like for
example in extragalactic clusters of galaxies due to structure
formation (e.g. Kang, Ryu \& Jones \cite{kang96}), active galactic
nuclei (AGN) jets (e.g. Begelman et al. \cite{bbr84}), supernova
remnants (e.g. van der Laan \cite{laan62}), star forming regions
(e.g. Bykov \& Fleishman \cite{bykov92}), pulsar winds (e.g. Rees
\& Gunn \cite{rees74}), galactic relativistic jets  (e.g. Zealey,
Dopita \& Malin \cite{zealey80}), or young star jets (e.g. Araudo et~al.
\cite{araudo07} and references therein).

In the case of AGNs, the interaction of jets with their environment allows the
study of the jet properties, like its matter content and energy budget, either
in the case of Fanaroff-Riley~II (FRII; e.g. Scheck et al. \cite{sch02}; Krause
\cite{kra05}) and Fanaroff-Riley~I (FRI; e.g. Perucho \& Mart\'{\i} \cite{pm07})
sources. The simulation of extragalactic jets and their interaction with the
environment have been addressed by a number of authors during the last decades
(e.g., Mart\'{\i} et al. \cite{mart97}; Komissarov \& Falle \cite{kofa98}; Aloy
et al. \cite{alo99}; Rosen et al. \cite{ro99}; Scheck et al. \cite{sch02};
Leismann et al. \cite{leis05}; Perucho et al. \cite{pe+05,pe+06}; and Mizuno et
al. \cite{mizu07}). In particular, the evolution of jets in homogeneous and
inhomogeneous ambient media, the influence of the internal energy on the
structure of the jets, the mass load of external material, the evolution of FRI
and FRII sources, and the growth of the Kelvin-Helmholtz instability in
relativistic flows have been thoroughly studied.

In the case of microquasars, X-ray binary systems with relativistic jets  (e.g. Mirabel \& Rodr\'iguez \cite{mirabel99}; Rib\'o
\cite{ribo05}), there are not many studies on the consequences of the jet/medium interaction (e.g. Vel\'azquez \&  Raga \cite{velazquez00}).
A common approach has been to compare with extragalactic jets to point out similarities between the two types of sources (e.g., G\'omez
\cite{gomez01}, Hardee \& Hughes \cite{hardee03}). It seems clear that these galactic outflows should also generate some kind of structure
in their termination region, although the characteristics of this region from the radiative and dynamical point of view are unclear (e.g.
Aharonian \& Atoyan \cite{aharonian98}; Heinz \cite{heinz02}; Heinz \& Sunyaev \cite{heinzS02}; Bosch-Ramon et al. \cite{bosch05}; Bordas et~al., in preparation), and only a
handful of sources show hints or evidences of jet/medium interaction (e.g. SS~433, Zealey et~al. \cite{zealey80}; XTE~J1550$-$564, Corbel
et~al. \cite{corbel02}; Cygnus~X-3, Heindl et al. \cite{heindl03};
Cygnus~X-1, Mart\'i et al. \cite{marti96}, 
Gallo et~al. \cite{gallo05}; H1743$-$322, Corbel et~al. \cite{corbel05}; Circinus X-1, Tudose et al.
\cite{tudose06}; LS~I~+61~303, Paredes et~al. \cite{paredes07}).

A microquasar jet can interact strongly not only with the interstellar medium.
In high-mass microquasars (HMMQ), the massive and hot primary star suffers
severe mass loss in the form of a supersonic wind that can embed the
jet\footnote{We focus on HMMQs due to the relevance of the stellar wind at the
spatial scales studied here. The presence of dense material and the stellar photon
field makes HMMQs even more interesting from the radiative point of view. In the
case of low mass systems, there might be accretion disk winds affecting jet
dynamics (e.g. Tsinganos et al. 2004), but the properties of the environment are not well known and the impact on the jet is less clear.}. 
This wind can have strong
impact on the jet dynamics at the binary system spatial scales. In fact, the jet
may be destroyed by the wind, preventing jet detection. In such scenario, shocks
would be likely to occur, leading to efficient particle acceleration and the
production of multiwavelength radiation. All this shows the importance of
studying the interaction between the jet and the stellar wind using
hydrodynamics. Previous works treated the jet/stellar wind interaction using a
more phenomenological approach and focused on possible radiative outcomes (e.g.
Romero et~al. \cite{romero03}, Romero et~al. \cite{romero05}, Aharonian et~al.
\cite{aharonian06a}). Regarding the production of non-thermal particles in the
present context, diffusive shock (first order Fermi) acceleration is a plausible
mechanism (e.g. Drury \cite{drury83}). Other mechanisms of particle acceleration
may be operating as well, like second order Fermi acceleration (Fermi
\cite{fermi49}), shear acceleration (e.g. Rieger \& Duffy \cite{rieger04}), or
the so-called converter mechanism (e.g. Derishev et al. \cite{derishev03}).

In this work, we present the results from 2-dimensional relativistic
hydrodynamical simulations, with cylindrical and slab symmetry, in a simplified
scenario of the jet/stellar wind interaction in an HMMQ. Our goal is mainly to
make first order estimates of the dynamical impact of the wind on the jet. In
addition to this, we provide with semi-analytical estimates on the radiative
outcomes.

In the next section, we briefly describe the main features of the scenario
considered and present the physical parameters of the simulations; in
Sect.~\ref{sim}, we characterize the setup of the simulations and explain their
technical aspects; in Sect.~\ref{res}, we present the results obtained, which
are discussed in Sect.~\ref{disc} together with the consequences derived from
the simulations from the points of view of the radiation or jet stability. We
summarize all this in Sect.~\ref{sum}.

\section{Physical scenario}\label{phys}

The physical scenario adopted here corresponds to an HMMQ with physical parameters similar to those of
Cygnus~X-1 or LS~5039, two close high-mass X-ray binaries with jets and with moderate-to-strong stellar mass
loss from a primary O star, in the form of a fast and supersonic wind. The distance between the jet base and
the primary star is fixed to $R_{\rm orb}=3\times 10^{12}$~cm. This distance is similar to those 
in LS~5039 (Casares et al.~\cite{cas05}) and Cygnus~X-1 (Gies \& Bolton~\cite{gb86}). 
Typical luminosities and photon energies for O stars are $L_*\approx 10^{39}$~erg~s$^{-1}$ and
$\epsilon_0\approx3\,kT\approx 10$~eV, respectively; the stellar wind present speeds $V_{\rm w}\approx
2\times 10^8$~cm~s$^{-1}$, mass loss rates $\dot{M_{\rm w}}\approx 10^{-6}$~M$_{\odot}$~yr$^{-1}$, and
temperatures $T_{\rm w}\sim 10^4$~K.

For the jet, we adopt a hydrodynamical supersonic outflow ejected
perpendicular to the orbital plane at mildly relativistic
velocities, $V_{\rm j,0}\sim 10^{10}\,\rm{cm\,s^{-1}}$, and
temperature $T_{\rm j}\approx 10^{9-10}$~K$\ll m_{\rm p}c^2/k_{\rm
B}$, where $m_{\rm p}$ is the proton mass, $c$ is the speed of
light and $k_{\rm B}$ is the Boltzmann constant. The choice of the
temperature is somewhat arbitrary since our assumption is that the
jet is supersonic. Nevertheless, this temperature plays a role and
its relevance is discussed in Sect.~\ref{disc}. In this work,
we focus on jets formed by protons and electrons, as expected in
case the jet is fed with particles from the accretion disk. Given
the adopted temperatures and jet speeds, the most reasonable
choice of adiabatic exponent is $\Gamma=5/3$.

If we considered a leptonic jet, the adopted temperatures would imply an
important increase of the internal energy of the particles and
thus would significantly reduce the Mach number, affecting the
assumption of a supersonic flow used here. Scheck et
al~\cite{sch02} have shown that a supersonic leptonic jet with the same
energy flux than a hadronic jet, develops in a very similar way.
Therefore, large differences are not expected if the kinetic
luminosity of the jet is kept constant regardless of the
composition.

Regarding the geometry adopted for the simulation, a cylindrical symmetry can suffice to study the jet head
shock formation in the wind. Since the jet speed $V_{\rm j}\gg V_{\rm w}$, the wind speed (which breaks the
cylindrical symmetry) is neglected. As it will be seen in the slab simulations, this assumption is
reasonable for powerful jets, but not in the case of weaker ones. At the present stage, slab symmetry
simulations are enough to show the importance of the stellar wind ram pressure on the jet, because the
timescales on which the wind surrounds the jet ($\ga R_{\rm j}/V_{\rm w}$; where $R_{\rm j}$ is the jet
radius) are longer than the simulation timescales. Also, the adopted configuration of the stellar wind,
coming from a fixed side of the jet, is consistent with an orbital timescale much longer than the simulation
timescales. The adopted simplifications are reasonable in our case, since we aim at pointing out remarkable
features and not at obtaining accurate quantitative predictions.

We note that the wind structure is a key point for our simulations. It has been proposed that massive
stellar winds may be porous (e.g. Owocki \& Cohen 2006). In addition, mass-loss rates of OB stars may be
smaller up to a factor of several due to wind clumping, as shown, e.g., in Puls et al. (2006), where estimates
for the real mass-loss rates for several O stars after wind clumpiness correction are given. For very
porous winds, the collision of the jet with very small and dense clumps (i.e. the dynamically relevant part
of the wind) would be rare, although these interactions could have interesting radiative outcomes  (e.g.
Aharonian \& Atoyan \cite{aharonian96}; Romero et~al. \cite{clumps}). Also, smaller mass loss rates would
weaken to some extent the jet kinetic luminosity constraints derived from this work.

In our simulations, the jet starts at a distance $z_0=6\times 10^{10}\,\rm{cm}$ from the compact object,
where it should be still little affected dynamically by the environment. The initial half-opening angle
of the jet is taken to be 0.1~radians, which is given by the relation between the advance and expansion
velocities\footnote{A jet in free expansion due to a large initial overpressure expands approximately at its sound
speed (\cite{lea91}), which is, in our simulations, roughly an order of magnitude smaller than the advance
speed (for the temperatures given in Sect.~\ref{sim}), which results in the opening angle given above.}.
Then, the jet propagates through the system sweeping up stellar wind material. Two shocks are eventually
generated, one in the stellar wind (i.e. forward or bow shock) and another one in the jet itself (i.e.
backwards or reverse shock). The shocked material is called here shell for the wind (i.e. bow shock
downstream) and cocoon for the jet (i.e. reverse shock downstream). The jet continues its advance slowing
down due to kinetic energy exchange with the swept up wind material, and eventually could be bent, and even
destroyed, by the lateral stellar wind ram pressure. Also, the pressure of the wind or the cocoon as compared to the
lateral pressure of the jet can give rise to strong recollimation shocks. Recollimation shocks may also
destroy the jet via deceleration and loss of collimation of the fluid downstream of these shocks (Perucho \&
Mart\'{\i} \cite{pm07}).

\section{Jet/wind interaction simulations}\label{sim}

 \begin{table}[]
  \begin{center}
  \caption[]{Parameters of the wind}
  \label{tab2}
  \begin{tabular}{ll}
  \hline\noalign{\smallskip}
  \hline\noalign{\smallskip}
Parameter & \\
  \hline\noalign{\smallskip}
Wind density (g~cm$^{-3}$) & $2.8\times 10^{-15}$ \\
Wind pressure (erg~cm$^{-3}$)& $1.5\times 10^{-3}$  \\
Wind velocity (cm~s$^{-1}$) &$2\times 10^8$\\
Wind mass-loss rate (M$_{\odot}$~yr$^{-1}$)&$10^{-6}$\\
Wind ram-pressure (erg~cm$^{-3}$) & $1.12 \times 10^2$\\
  \noalign{\smallskip}\hline
  \end{tabular}
  \end{center}
\end{table}

 \begin{table}[]
  \begin{center}
  \caption[]{Parameters of the jets}
  \label{tab1}
  \begin{tabular}{llll}
  \hline\noalign{\smallskip}
  \hline\noalign{\smallskip}
Parameter & I & II & III \\
  \hline\noalign{\smallskip}
Jet power (erg~s$^{-1}$)& $3.0\times 10^{34}$ & $10^{36}$& $3.0\times 10^{37}$\\
Jet init. pressure (erg~cm$^{-3}$)& $9.1$ & $68$& $6.2\times 10^4$\\
Jet init. density (g~cm$^{-3}$)& $2.2\times 10^{-16}$ & $5.9\times 10^{-16}$& $1.8\times 10^{-14}$\\
Jet init. temperature (K)
&$7.4\times 10^{8}$&$2.1 \times 10^{9}$&$6.2 \times 10^{10}$\\
Jet init. speed (cm~s$^{-1}$) & $1.3\times 10^{10}$ & $2.2\times 10^{10}$ & $2.2\times 10^{10}$\\
Jet init. Mach number &47.02&51.05&9.35\\
  \noalign{\smallskip}\hline
  \end{tabular}
  \end{center}
\end{table}

We have performed five numerical simulations, three with cylindrical
symmetry and two with slab symmetry, in order to study the evolution
of HMMQ jets with different injection power.

The simulations were performed using a 2-dimensional finite-difference code based on a high-resolution
shock-capturing scheme which solves the equations of relativistic hydrodynamics written in conservation
form. This code is an upgrade of the code described in Mart\'{\i} et al. (\cite{mart97}) (e.g. see Perucho et
al. \cite{pe+05}). Simulations were performed in two dual-core processors in the Max-Planck-Institut f\"ur
Radioastronomie.

The numerical grid of the cylindrical geometry simulations is formed by 320 cells in the radial direction
and 2400 cells in the axial direction in an uniform region, with physical dimensions of 20$\times\,300$
$R_{\rm j}$. An expanded grid with 160 cells in the transversal direction brings the boundary from
$20\,R_{\rm j}$ to $60\,R_{\rm j}$, whereas an extended grid in the axial direction composed by 480 extra
cells spans the grid axially from $300\,R_{\rm j}$ to $450\,R_{\rm j}$. The enlargement of the grid is done
to take the boundary conditions far enough from the region of interest and avoid numerical reflection of
waves in the boundaries affecting our results. The numerical resolution in the uniform grid is thus of 16
cells/$R_{\rm j}$ in the radial direction and 8 cells/$R_{\rm j}$ in the axial direction. Outflow boundary
conditions are used on the outer boundaries of the grid, inflow at injection, and reflection at the jet axis
in the cylindrical case. In the simulations, all the physical variables are scaled to the units of the
code, which are the jet radius, $R_{\rm j}$, the rest-mass density of the ambient medium, and the speed of
light.

The jet is injected in the grid at a distance of $6\times 10^{10} \rm{cm}$ from the compact object, and its
initial radius is taken to be $R_{\rm j,0}=6\times 10^{9} \rm{cm}$. The time unit of the code is thus
equivalent to 0.2~seconds, as derived from the radius of the jet at injection and the speed of light
($R_{\rm j,0}/c$). The grid covers the distance between $6\times 10^{10}\,\rm{cm}$ and $2\times
10^{12}\,\rm{cm}$, i.e., a 60\% of $R_{\rm orb}$. The ambient medium, i.e. the stellar wind, is composed by
a gas with thermodynamical properties derived from Sect.~\ref{phys} (see Table~\ref{tab2}). Both the jet and
the ambient medium are considered to be formed by a non relativistic gas with adiabatic exponent
$\Gamma=5/3$.

In the simulations the jets are injected with different
properties. The physical parameters that characterize each
simulation are listed in Table~\ref{tab1}, the main difference
between the three cases being the kinetic luminosity ($L_{\rm
j}$): weak jet (case I) of $3\times 10^{34}$~erg~s$^{-1}$; mild
jet (case II) of $10^{36}$~erg~s$^{-1}$; powerful jet (case III)
of $3\times 10^{37}$~erg~s$^{-1}$. The selection of the jet
power is addressed to show some illustrative cases: the lowest
value, that of case I, is similar to the minimum power required to
power radio emission under reasonable assumptions of the radio
emitter (e.g. for Cygnus~X-1, see Heinz~\cite{heinz06}); that of
case II is a bit smaller than the power estimates found for
Cygnus~X-1 (Gallo et al.~\cite{gallo05}), and similar to the
$L_{\rm j}$ inferred for LS~5039 (Paredes et
al.~\cite{paredes06}), and the maximum value, that of case III, is
between the upper limit for the Cygnus~X-1 $L_{\rm j}$  and the
$L_{\rm j}$ of SS~433 (e.g. Gallo et al.~\cite{gallo05}; Marshall
et al.~\cite{marshall02}). These jets are evolved until they
have reached distances similar to the binary system size.

The velocities of the jets were selected as those of mildly
relativistic flows, with Lorentz factors $\gamma_{\rm j}=1.1$ for
case I, and $\gamma_{\rm j}=1.5$ for cases II and III. The jet Mach
numbers are similar in cases I (47.02) and II (51.05), whereas it
is much smaller in case III (9.35). In the latter, the increase in
the jet power is due to an increase in the internal energy of the
jet, keeping the same jet speed as in case II. The jet
temperatures also show this fact, as the jet in case III has a
temperature larger than that of case II by a factor 30. This
allows us to see the influence of changes of velocity and
internal energy on the jet evolution.

In the case of the slab geometry simulations, the grid size is shortened to $200\,R_{\rm j}$ along the jet
axis, and it is doubled in the direction transversal to the jet axis (it extends from $-60\,R_{\rm j}$ to
$60\,R_{\rm j}$). The numerical resolution is the same as that used in the cylindrical simulations. The jet
is injected in an ambient medium that mimics the presence of a spherical wind centered in a star at a
distance $R_{\rm orb}$ from the compact object on the orbital plane, with the properties mentioned in
Sect.~\ref{phys}. In this case, the boundary conditions are injection on the side from which the wind is
assumed to come and on the base of the jet, and outflow in the outer boundaries on the opposite side of the
grid and at the end of the grid in the axial direction. This kind of simulations were done for cases I and
II\footnote{The slab {\it jets} have the same lateral pressure and $L_{\rm j}/{\rm cm}$ -on the simulation
plane- as in the cylindrical case.}, since the interaction with the wind was expected to be possibly
relevant for the jet dynamics. 

\section{Results}\label{res}

\subsection{Case I: a weak jet}

{\bf Cylindrical jet}

The jet in case I needs $t_{\rm f}\sim900\, \rm{s}$ to cover the distance between the injection and the
outer boundary of the numerical grid. The bow shock generated by the injection of the jet in the ambient
medium moves at a mean speed of $V_{\rm bs}\sim 0.06\,c$. The jet generates a high pressure cocoon that keeps
it collimated (see the discussion in the next section). No other shocks or significant jet perturbations are
generated, and the jet propagates well collimated through the wind. We note that the speed of the unshocked
jet does not change significantly along the axis. We do not show any image of this simulation due to the
lack of remarkable features. A similar result has been obtained in case II, we thus refer the reader to
Fig.~\ref{fig:mild} (see Sect.~\ref{sec:mild}).

The approximation to a {\it static wind} is reviewed critically in view of the results of the slab jet
simulation presented next, in which the wind has the appropriate velocity.

{\bf Slab jet}

\begin{figure*}[!t]
\centerline{
\psfig{file=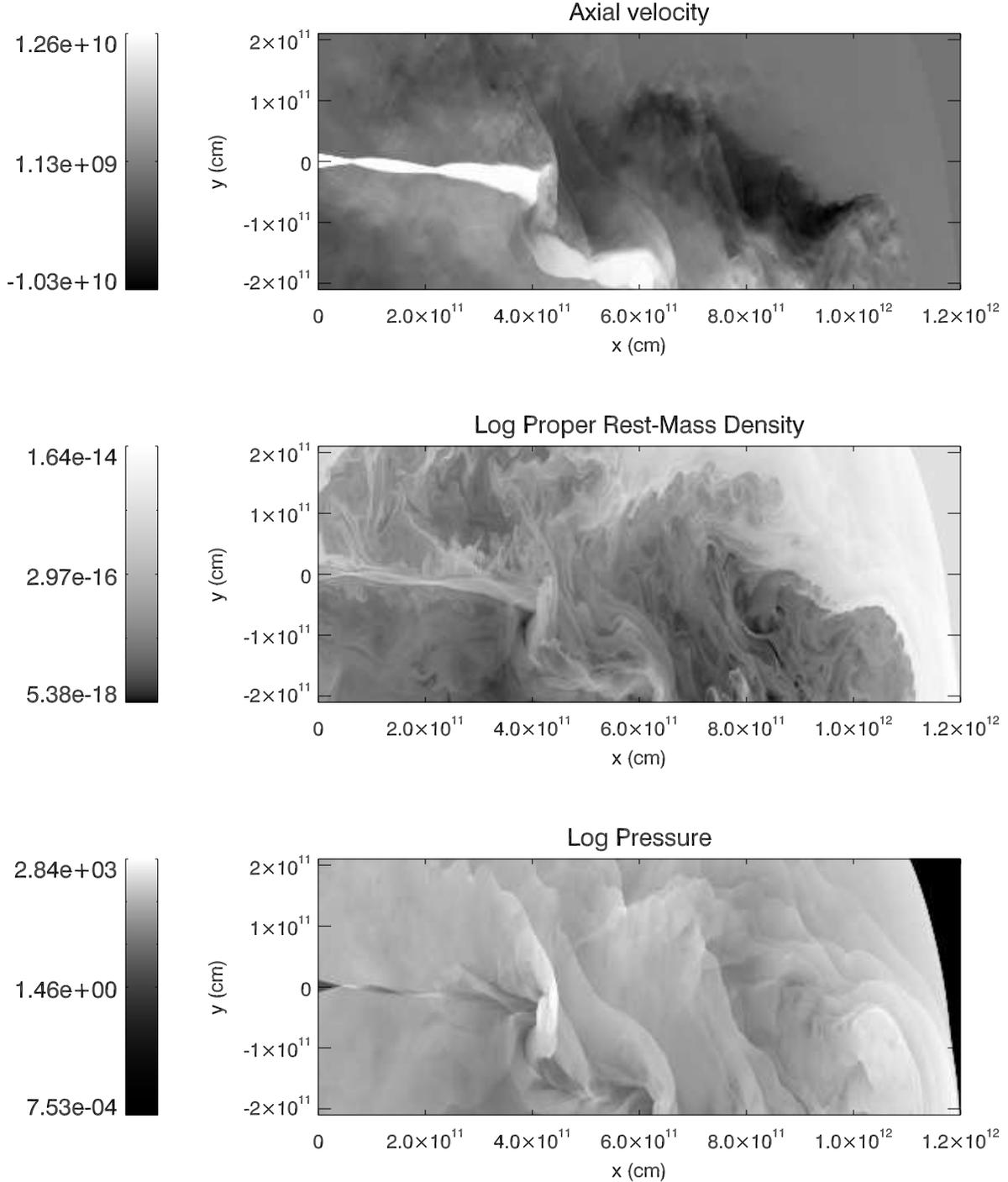,width=\textwidth,angle=0,clip=}}
\caption{Maps of axial velocity (cm~s$^{-1}$), rest mass density
(g~cm$^{-3}$) and pressure (dyn) at the end of the simulation
($t_{\rm f}=680\,\rm{s}$) of a weak jet in slab geometry, case I.
The horizontal and vertical coordinates indicate distances (cm) to
the injection point in the numerical grid and the jet axis, respectively.
The stellar wind is centered on the position of the
primary star, at $x=-6\times 10^{10}$~cm and $y=3\times 10^{12}$~cm in
the figure.} \label{fig:weak}
\end{figure*}

The same simulation has been performed in planar coordinates including the wind motion. The wind comes from
a point source at the distance and position of the primary star. Figure~\ref{fig:weak} shows panels of
axial velocity, rest mass density and pressure for the moment when the jet head has left the grid. The
impact of the wind in the evolution of weak jets is clear from the plots. The bow shock generated by the jet
moves at $0.05\,c$ and is deformed by the wind and the asymmetry in pressure on both sides of the jet. The
wind also pushes the jet sideways, and two diagonal shocks at $z\la 3\times 10^{11}\,\rm{cm}$ are formed due
to the interaction with the environment, which is in overpressure with respect to the jet flow. After the
first shock, the jet widens and recollimates, starting a new process of expansion after $z\approx 2.6\times
10^{11}\,\rm{cm}$. The jet is finally disrupted due to the action of the wind, after the jet has lost some
of its initial inertia due to the previous internal shocks. A strong shock is generated in the disruption
point at $z\approx 5\times 10^{11}\,\rm{cm}$. Farther from this point the jet is significantly deviated and
a portion of fast material of the remaining outflow is observed to leave the regular numerical grid sideways
at $z\sim 6-7\times 10^{11}\,\rm{cm}$. This simulation shows an example of a frustrated jet turned into a
decollimated, slow outflow due to the effect of the wind.

\subsection{Case II: a mild jet}

{\bf Cylindrical jet} \label{sec:mild}

\begin{figure*}[!t]
\centerline{
\psfig{file=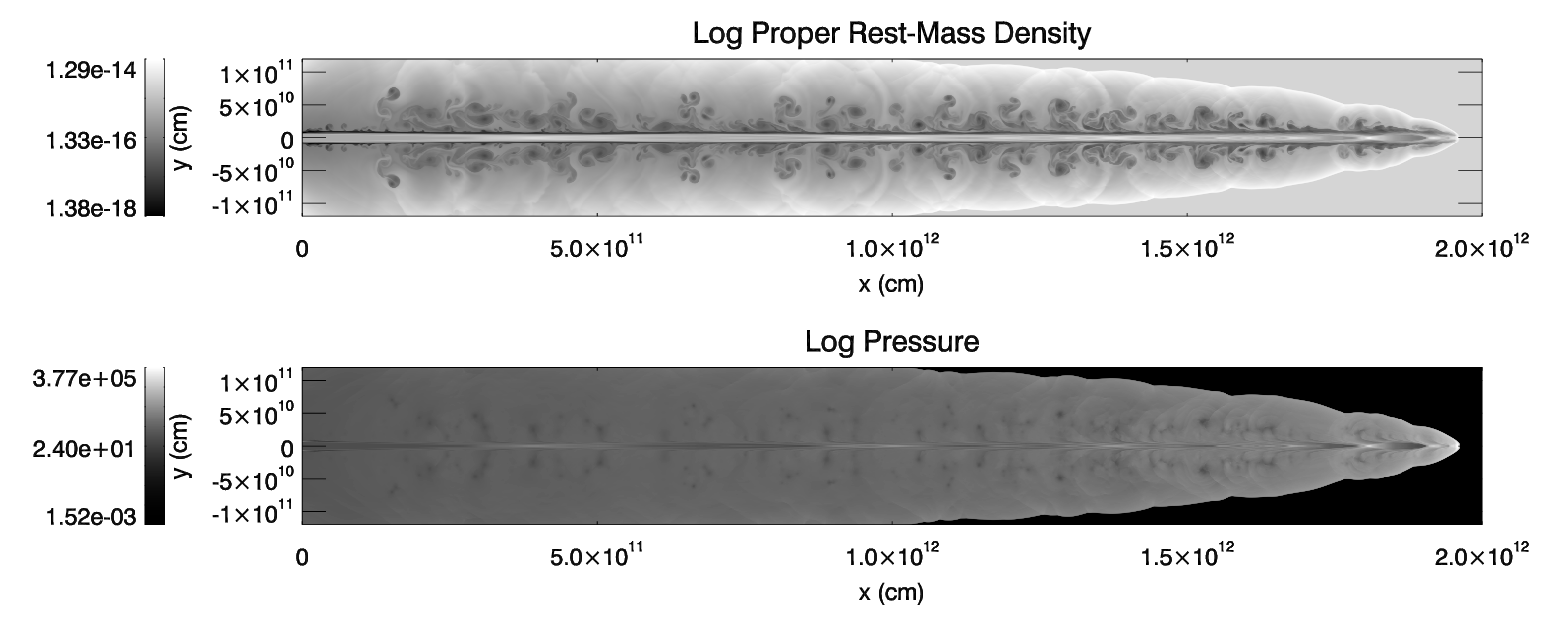,width=\textwidth,angle=0,clip=}}
\caption{Maps of rest mass density (g~cm$^{-3}$) and pressure
(dyn) at the end of the simulation ($t_{\rm f}=164\,\rm{s}$) of
the mild jet (case II) in cylindric geometry. The horizontal and
vertical coordinates, and the star location, are like in Fig.~\ref{fig:weak}.}
\label{fig:mild}
\end{figure*}

The simulation of case II in cylindrical coordinates is very similar to that of case I. Fig.~\ref{fig:mild}
shows the rest-mass density and pressure of the jet when its head has reached the end of the grid. The main
difference between this case and case I lies on the fact that here some internal shocks are produced by the
slight overpressure of the jet with respect to the cocoon at $z> 1.4\times 10^{12}\,\rm{cm}$. Before, the
jet is efficiently collimated by the high pressure cocoon. The bow-shock propagates at $V_{\rm bs}\sim
0.3\,c$, so it is significantly faster than in the previous case. The bow-shock introduces a jump in
pressure of $10^8$ with respect to the cold wind. The internal, conical shocks observed in the jet itself,
show jumps in pressure of $10^2-10^4$, although there are no strong changes of axial velocity with respect
to the unshocked jet material. This implies that not much kinetic energy could be transferred to non-thermal 
particles via these shocks.

{\bf Slab jet}

\begin{figure*}[!t]
\centerline{
\psfig{file=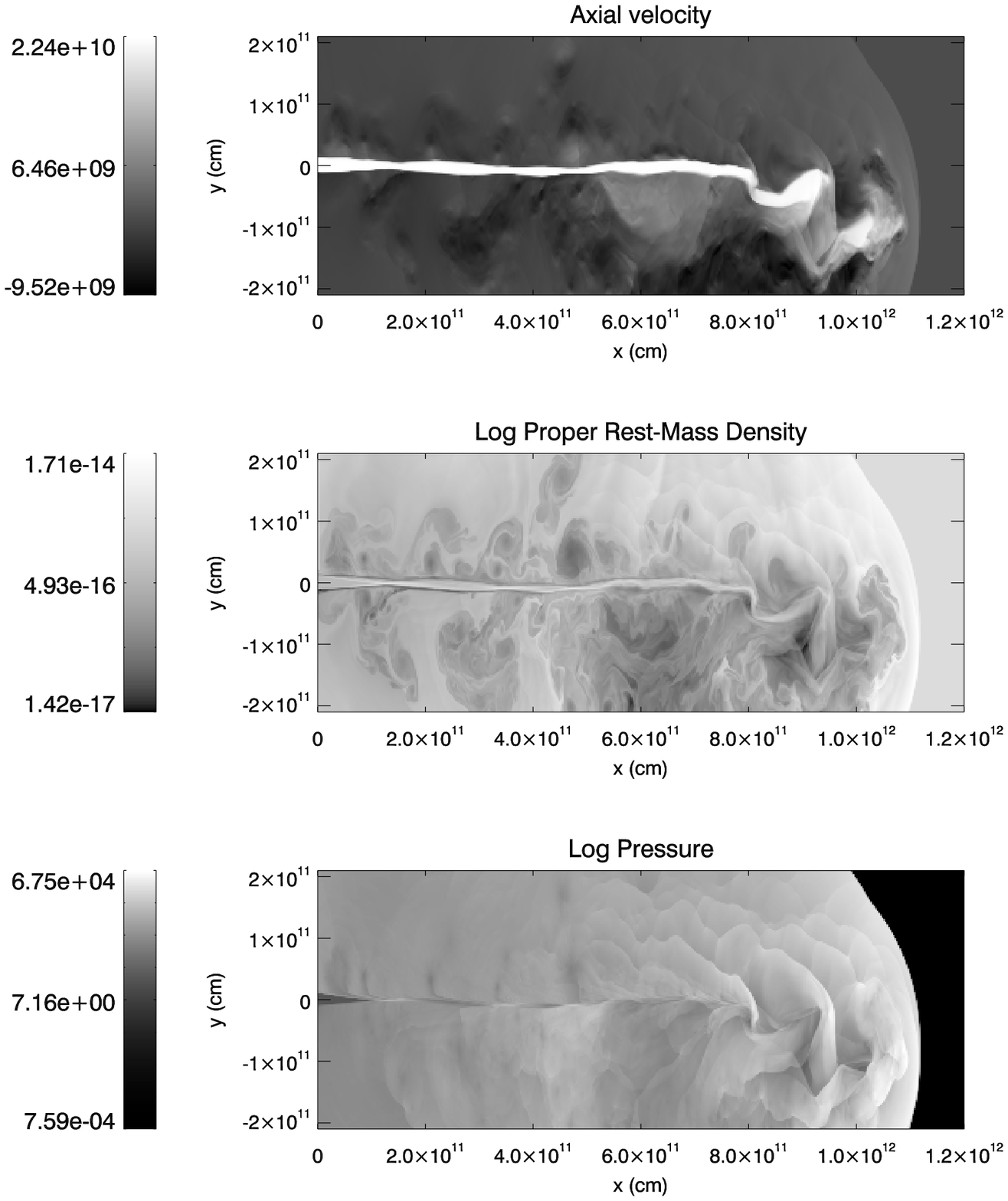,width=\textwidth,angle=0,clip=}}
\caption{Maps of axial velocity (cm~s$^{-1}$), rest mass density
(g~cm$^{-3}$) and pressure (dyn) at the end of the simulation
($t_{\rm f}=140\,\rm{s}$) of a mild jet in slab geometry, case II.
The horizontal and
vertical coordinates, and the star location, are like in Fig.~\ref{fig:weak}.}
\label{fig:mild-slab}
\end{figure*}

Also in case II, a slab simulation was performed accounting for the velocity of the stellar wind from the
primary star. Fig.~\ref{fig:mild-slab} shows several maps of physical magnitudes from the simulation. The
bow-shock is also faster than that in the slab case I ($V_{\rm bs}\sim 0.3\,c$). Several diagonal shocks are
formed due to the interaction with the wind along the jet trajectory. Again, the jet is in underpressure
with respect to the ambient, what generates an initial collimation. The pinching triggered by the first
collimation shock and subsequent expansion of the jet, together with the thrust exerted by the wind on the
jet, generate nonlinear structures on the latter. In this case, the jet is also finally disrupted at $z\sim
10^{12}\,\rm{cm}$ (see Fig.~\ref{fig:mild-slab}). The pattern of shocks at the disruption region is similar
to that observed in Fig.~\ref{fig:weak}. These results show that the jet can be disrupted by the
stellar wind ram pressure even for intermediate kinetic luminosities.

\subsection{Case III: a powerful jet}

\begin{figure*}[!t]
\centerline{
\psfig{file=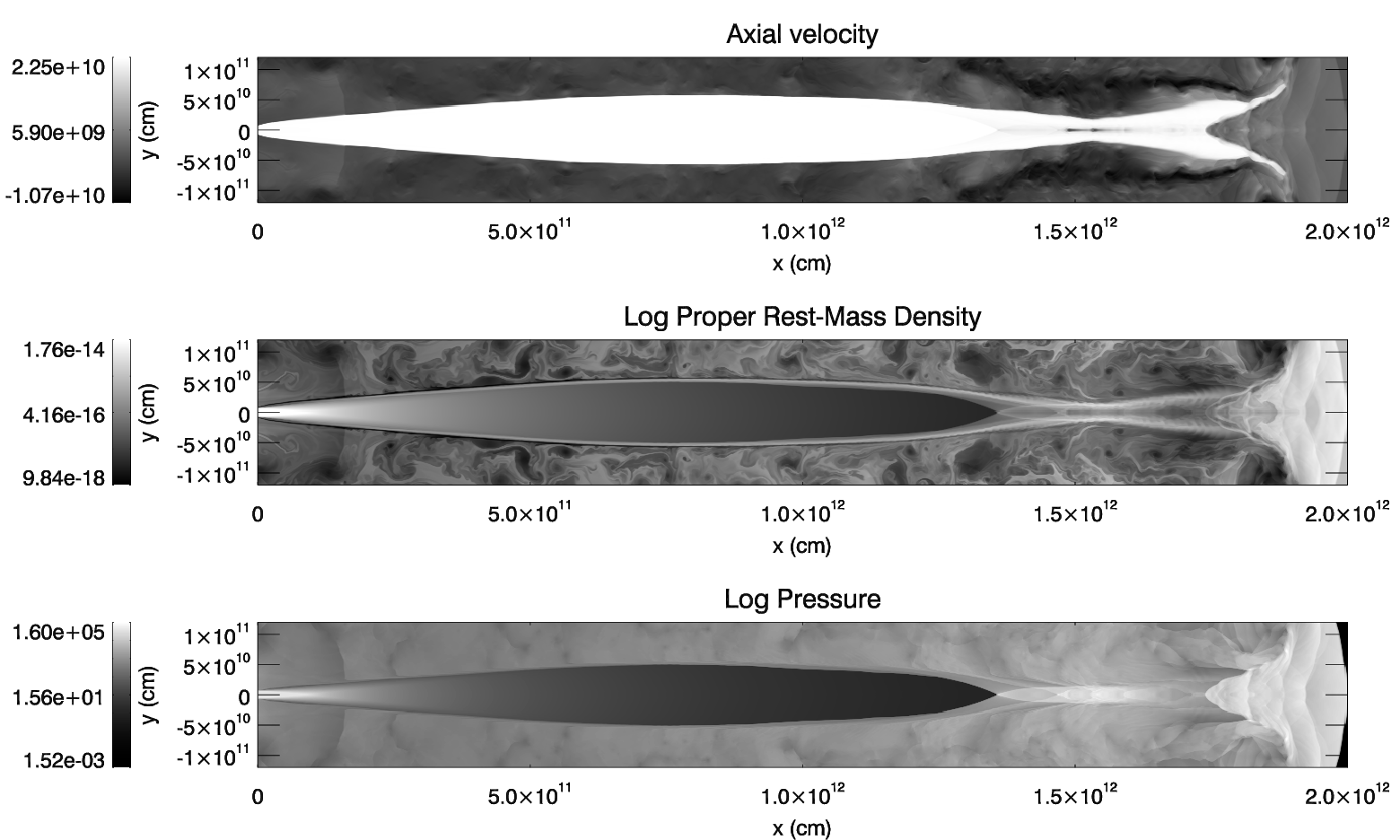,width=\textwidth,angle=0,clip=}}
\caption{Maps of axial velocity (cm/s), rest mass density
(g~cm$^{-3}$) and pressure (dyn) at the end of the simulation
($t_{\rm f}=212\,\rm{s}$), for a powerful jet (case III). The horizontal and
vertical coordinates, and the star location, are like in Fig.~\ref{fig:weak}.}
\label{fig:pow}
\end{figure*}

The simulation of the powerful jet ($L_{\rm j}=3\times 10^{37}\,\rm{erg\,s^{-1}}$) shows important
differences with respect to the previous cylindrical simulations. Fig.~\ref{fig:pow} displays the maps of
rest-mass density, pressure and axial velocity when the bow-shock is about to reach the limit of the regular
grid. The bow-shock propagates at $V_{\rm bs}\sim 0.3\,c$. This $V_{\rm bs}$ is the same as in case II
because part of the larger initial thrust of case III jet is invested in lateral expansion and 
heating. The latter happens due to, in contrast to
the previous cases, a strong recollimation shock formed at $z\approx 1.3\times 10^{12}\,\rm{cm}$. This
shock is formed due to the initial overpressure of the jet with respect to the cocoon. 

The relation between the pressure in the cocoon and that in the jet can make possible the formation of such
shocks. The jump in pressure generated in the jet axis is of the order of $10^4$. In a second stage of the
jet evolution, i.e., once the bow-shock is away from the studied region and the cocoon is substituted by the
stellar wind, recollimation shocks may change their position and properties. However, as is shown in the next
section, these shocks may still be present. Such shocks, if strong enough, can trigger pinching of the jet, as
observed in Fig.~\ref{fig:pow}, which, in turn, may end up in mass load of ambient gas and jet disruption
(Perucho \& Mart\'{\i} \cite{pm07}). Hence the importance of these recollimation shocks, not
only from the point of view of particle acceleration and subsequent radiation, but also from the dynamics of
the jet.

\section{Discussion}\label{disc}

\subsection{On the formation of shocks in the studied
region}\label{sectrecolanalit}

We see that two different kinds of shock, important for the acceleration of particles and subsequent
radiation, may form: reverse/bow shocks, which cross the region of constant density (within $\sim R_{\rm
orb}$) at a fraction of $c$ on timescales $<10^4\,\rm{s}$; and recollimation shocks, generated by the
interaction of the jet with its environment, the cocoon or the stellar wind itself. Unlike the reverse/bow
shock, which appears naturally as shown by the simulations, the formation of recollimation shocks
depends on the complex jet/medium pressure balance and deserves further analysis. The possibility that they
arise is discussed in this section in terms of the properties of the jets and their environments.

Recollimation shocks form when an initially overpressured jet expands and reaches a lower pressure than that
of its environment. In order to know whether a jet generates such a shock, and at which distance, the change
of jet pressure with distance is needed to be compared with that of the ambient medium. In an adiabatic
expansion, the condition $\rho_{\rm j}\gamma_{\rm j} V_{\rm j} A_{\rm j} = C_1$ holds, where $C_1$ is a
constant and $A_{\rm j}$ is the jet cross section. In our simulations, the jet velocity and Lorentz factor
are basically constant, implying $\rho_{\rm j}\propto 1/R_{\rm j}^2$, which, for a conical jet, transforms
into $\rho_{\rm j}\propto 1/z^2$, i.e., the density of the jet decreases as the square of the distance to
the source. It is also known that the jet pressure ($P_{\rm j}$) and density satisfy the relation $P_{\rm
j}/\rho_{\rm j}^{\Gamma}=C_2$, where $\Gamma$ is the adiabatic exponent and $C_2$ a constant. Thus, 
$P_{\rm j}\propto 1/z^{2\Gamma}$, which implies ($\Gamma=5/3$):
\begin{equation}
\label{eq:pres}
P\propto z^{-10/3}.
\end{equation}
A minimum distance for the position of a recollimation shock can be given as the place where the pressure in
the jet is the same as the pressure of the ambient ($z_{\rm eq}$). \footnote{Actually, the shock occurs at a
distance given by $z_{\rm eq}$, the Mach number of the jet at this point ($M$), and the jet radius at $z_{\rm
eq}$ ($R_{\rm eq}$): $z\sim z_{\rm eq}+R_{\rm eq}\times M$ (see \cite{lea91}), as the information takes a
certain time to reach the jet axis. At this distance, the pressure in the jet is strictly smaller than that of
the ambient. In fact, the distance at which the curvature of the jet starts to change, which can be
identified with the place where $P_{\rm j}\sim P_{\rm ext}$, is about half of that in which the shock occurs
(see Fig.~\ref{fig:pow}). The second part of the equation depends on the parameters of the jet  in the
equilibrium point. For simplicity, we adopt the $z_{\rm eq}$ as a lower-limit approximation to the position
of the shock.} We can express this distance in terms of the pressure of the jet at a reference position
$z_0$ ($P_{\rm j,0}$), which will be taken as the injection point in the grid, and the pressure of the
ambient medium ($P_{\rm ext}$). From Eq.~(\ref{eq:pres}):
\begin{equation}\label{eq:zxoc}
z_{\rm s}> \left(\frac{P_{\rm j,0}}{P_{\rm ext}}\right)^{3/10}
z_0.
\end{equation}
From this, certain limits on the pressure (and temperature) of
jets, suitable to develop recollimation shocks, can be given.
Since we are interested in the region of maximal wind/jet
interaction, we impose that the right hand side of the inequality
in Eq.~(\ref{eq:zxoc}) is $\leq 10^{12}\,\rm{cm}$, being a safe
estimate for $z_{\rm s}<R_{\rm orb}$. Thus, fixing $z_0$, an
estimate of the initial overpressure in the jet that may result in
a recollimation shock in $<R_{\rm orb}$ can be given. If $z_0$ is
taken as the injection point, i.e., $z_0\,=\,6\times
10^{10}\,\rm{cm}$, we obtain:
\begin{equation}\label{prj}
1\,<\,\frac{P_{\rm j,0}}{P_{\rm ext}}\,\leq\,1.2\times 10^4.
\end{equation}
Eq.~(\ref{prj}) can be written in terms of the jet temperature
($T_{\rm j,0}$) and the jet density ($\rho_{\rm j,0}$) at the
injection point:
\begin{equation}\label{jtemp}
T_{\rm j,0}\leq \frac{1.2\times 10^4\,P_{\rm
ext}\,m_p}{(\Gamma-1)\,\rho_{\rm j,0}\, k_{\rm
B}}=\frac{2.18\times 10^{-4}\,P_{\rm ext}}{\rho_{\rm
j,0}}\,\rm{K}.
\end{equation}

Both cylindrical and slab simulations show that, initially, the jets are embedded in their cocoons. However,
as shown by the slab jet simulations, quickly the stellar wind starts to dynamically affect the bow-shock
and the jet itself, and once the jet head has left the binary system, shocked wind will substitute cocoon
material. We note that the properties of the wind change via shocking with the jet. In this case, the
relative pressure of the jet with respect to the ambient would also change, at least on the side where the
wind impacts. This asymmetry in the ambient medium can give rise to asymmetric shocks, generated on one side
of the jet and propagating diagonally through it, as observed in the slab simulations (Figs.~\ref{fig:weak}
and \ref{fig:mild-slab}). In short, these two different scenarios, i.e. interaction with the cocoon and
the wind, have to be taken into account.

When the pressure of the ambient medium corresponds to that of the
cocoon, the following equation (Begelman \& Cioffi \cite{bc89})
estimates the pressure of the latter ($P_{\rm c}$) as a function
of the jet kinetic power, advance speed of the bow shock, and the
surface of interaction between the bow shock and the ambient medium
($A_c$):
\begin{equation}\label{pcoc}
P_{\rm c} \sim \frac{L_{\rm j}}{V_{\rm bs} A_{\rm c}}.
\end{equation}
When the external pressure is exerted by the wind on the jet, it
can be computed by assuming that the density of the shocked wind
is four times the original density of the wind, basing on the
Rankine-Hugoniot jump conditions, and that most of the kinetic
energy of the wind is thermalized and, thus, its internal energy
is similar to the original kinetic one.

{\bf Jet/Cocoon interaction}

For a typical jet with kinetic power $10^{36}\rm{erg\,s^{-1}}$,
bow-shock advance speed $0.1\,c$ and bow-shock radius around
$3\times 10^{11}\rm{cm}$, the cocoon pressure is $P_{\rm c}\sim
10^3\rm{erg\,cm^{-3}}$. Then, if the density in the jet at
injection ($z_0=6\times 10^{10}\rm{cm}$) is $\rho_{\rm j,0}\sim
10^{-15}\,\rm{g\,cm^{-3}}$, the temperature must be:
\begin{eqnarray}\label{tmax}
T_{\rm j,0}< 3\times 10^{14}\cdot\left(\frac{L_{\rm
j}}{10^{36}\,\rm{erg\, s^{-1}}}\right)\cdot \left(\frac{3\times
10^9\,\rm{cm\,s^{-1}}}{V_{\rm bs}}\right)\cdot
\nonumber {}\\
\qquad \qquad \qquad \qquad \left(\frac{3\times
10^{11}\,\rm{cm}}{R_{\rm c}}\right)^2\cdot
\left(\frac{10^{-15}\,\rm{g\,cm^{-3}}}{\rho_{\rm
j,0}}\right)\,\rm{K},
\end{eqnarray}
which is fulfilled accounting for the fact that the jet must be
supersonic (i.e. $T_{\rm j,0}<(\Gamma-1)m_{\rm p}c^2/k_{\rm B}\sim
10^{12}$~K).

One could also impose that the pressure at $z_0$ must be larger
than that of the ambient to have an initial expansion of the jet
(see Eq.~\ref{prj}). A lower limit in the temperature of the jet
may be set: $T_{\rm j,0}>T_{\rm j,0,max}/(1.2\times
10^4)\,\rm{K}$. In the cylindrical simulations of cases I and II
the temperatures of the jets are below this limit (see
Table~\ref{tab1}), so they do not generate recollimation shocks
(see Fig.~\ref{fig:mild}). Nevertheless, we note that the location
of the injection point in the numerical grid, $z_0$, is somewhat
arbitrary. As the pressure in the jets increases towards the
compact object like $z^{5\Gamma/3}$, it can be inferred that, with the
physical conditions derived from the simulations, the pressure in
the jet would, at some point, become larger than in the cocoon.
Hence, the appropriate conditions for the generation of a
recollimation shock in those simulations would then exist.

{\bf Jet/wind interaction}

In the case of the jet interacting with the stellar wind, the
pressure is derived as explained above to obtain $P_{\rm w}\sim
10^2\,\rm{erg\,cm^{-3}}$. Using the jet density given above for
the jet,
\begin{eqnarray}\label{tmax2}
T_{\rm j,0}< 3\times 10^{13}\cdot\left(\frac{\rho_{\rm w}}{2.8\times
10^{-15}\, \rm{g\,cm^{-3}}}\right) \cdot
\left(\frac{V_{\rm w}}{2\times 10^8\,\rm{cm\,s^{-1}}}\right)^2\cdot
\nonumber {}\\
\qquad \qquad \qquad \qquad
\left(\frac{10^{-15}\,\rm{g\,cm^{-3}}}{\rho_{\rm
j,0}}\right)\,\rm{K},
\end{eqnarray}
which is again fulfilled accounting for the fact that the jet must
be supersonic.

We conclude that the presence of standing shocks within the binary
system is very probable within our prediction limits.

\subsection{Radiative processes}

As discussed above, strong shocks take place in the jet head, i.e. the bow/reverse shocks, and in the
jet/cocoon contact surface, i.e. the recollimation shocks. In practice, the capability of the jet head structures to
produce non-thermal emission is linked to the time they remain within the system, since the shock energy
transfer reduces further out with the strong decrease of the wind density. In addition, quasi-permanent
recollimation shocks between the jet and the stellar wind can also occur, as noted in
Sect.~\ref{sectrecolanalit}. Since all these shocks are strong, and we know approximately the speed of the
shocks from the simulations, we can estimate the particle acceleration efficiency. In the following, we
compute the maximum energy and the emission spectral energy distribution of particles accelerated
under shock conditions similar to those shown by the simulations.

\subsubsection{Non-thermal particle production}

{\bf Particle acceleration}

As a first order approach, and provided that the shock Lorentz
factors are $\sim 1$, we compute the particle acceleration rate in
the context of the non-relativistic (first order Fermi) diffusive
shock acceleration mechanism, in the test particle approximation,
1-dimensional case and with dynamically negligible magnetic field.
This mechanism typically predicts, for strong shocks, a power-law particle
distribution of index -2 (e.g. Drury \cite{drury83}). The
acceleration rate in cgs units can be computed using:
\begin{equation}
\dot{E}=\eta qBc \label{acc},
\end{equation}
where $\dot{E}$ is the particle energy gain per time unit, $q$ the
particle charge, $B$ the magnetic field in the accelerator and
$\eta$ the acceleration efficiency coefficient, which can be
calculated through the formula (e.g. Protheroe \cite{protheroe99}):
\begin{equation}
\eta=\frac{3}{20K_{\rm d}}\left(\frac{V_{\rm S}}{c}\right)^2,
\end{equation}
for a parallel magnetic field, where $K_{\rm d}$ is proportional
to the diffusion coefficient ($K_{\rm d}=1$ corresponds to the
Bohm regime) and $V_{\rm S}$ is the speed of the shock in the
upstream reference frame. The value of the second parameter can be
obtained from the simulations, allowing to estimate $\eta$. For
the reverse and forward shocks, we have taken typical velocities
when the jet head is still within the binary system, obtaining
$V_{\rm S}\sim 10^{10}$~cm~s$^{-1}$. In case of recollimation
shocks, as shown in Fig.~\ref{fig:pow}, velocity jumps could reach
similar values. In the Bohm diffusion case,
$\eta\sim 2\times 10^{-2}$. 
We recall that the recollimation shock shown in Fig.~\ref{fig:pow}
is related to the cocoon, although such structure could also be produced by
interaction of the jet lateral surface with the stellar wind.

Concerning the $B$ strength, the hydrodynamical assumption
imposes an energy density for the adopted magnetic field
$\ll$ the (internal plus kinetic) matter energy density.
Thus, the
value of $B$ is to be well below equipartition with jet matter
to be consistent with our simulations. For the
(internal plus kinetic) matter energy density in the shocked material given by the simulations, which
ranges $\sim 10^2-10^6$~erg~cm$^{-3}$, $B$ should be $\ll
10^2-10^4$~G depending on the jet power and shock properties.

We can compute the maximum energy of
the particles equating Eq.~(\ref{acc}) to the particle energy
losses. 

For electrons, when the
synchrotron energy losses are dominant, the maximum
energy can be obtained from:
\begin{equation}
E_{\rm max}  \approx 50 \sqrt{\frac{\eta}{B}}~{\rm TeV}.
\end{equation}
For inverse Compton (IC) losses in the Thomson regime ($E\ll 5.1\times 10^5~{\rm
eV}/\epsilon_0$; $\epsilon_0=10~{\rm eV}$), and accounting for the
stellar radiation density in the jet, $U_*\approx
300$~erg~cm$^{-3}$, the maximum particle energy is:
\begin{equation}
E_{\rm max}  \approx \sqrt{B\eta}~{\rm TeV},
\end{equation}
which is to be substituted by the following formula in the
Klein-Nishina (KN) regime ($E> 5.1\times 10^5~{\rm
eV}/\epsilon_0$) (Khangulyan, Aharonian \& Bosch-Ramon \cite{khangulyan07}):
\begin{equation}
E_{\rm max}  \approx 10^9(B\eta)^{3.3}~{\rm TeV}.
\end{equation}

$E_{\rm max}$ is determined by the smallest among the shown $E_{\rm max}$. For $K_{\rm d}=1$, $\eta=2\times
10^{-2}$, synchrotron energy losses are dominant for $B\ga 0.2$~G. $E_{\rm max}$ can reach 1~TeV for
$B=50$~G, and 10~TeV for $B=0.4$~G.

In the case of inefficient energy losses, as it commonly happens for protons, the size
of the accelerator, approximated here as the diameter of the jet
at the shock location ($D_{\rm j,S}$), gives the very upper-limit
to the maximum particle energy when equated to the particle mean
free path (the gyroradius in our case):
\begin{equation}
D_{\rm j,S}=r_{\rm g}=\frac{E_{\rm max}}{qB}.
\end{equation}
Higher energy particles escape the accelerator since they cannot
be confined by the magnetic field. For $D_{\rm j,S}\sim
10^{11}$~cm, $E_{\rm max}$ may reach $\sim 10^3$~TeV. 

{\bf Characterizing the particle energy distribution}

Once electrons are injected in the emitter, they suffer losses via synchrotron and IC processes.
Relativistic Bremsstrahlung and ionization losses of electrons are negligible due to the long timescales and
the high ionization degree of the shocked material, respectively. Protons lose energy mainly via
proton-proton collisions, although the efficiency of this process is in general small. All the particles are
convected with timescales that are  
shorter than the radiative ones for the electrons with the higher energies, and
protons of any energy. The convection timescale is approximated by $\tau_{\rm conv}\sim l_{\rm d}/V_{\rm
d}\approx 400$~s, being $l_{\rm d}\approx 10^{12}$~cm and $V_{\rm d}\approx V_{\rm S}/4$ (from the strong
shock jump conditions) the length of the downstream region and the convection velocity, respectively
(derived from the simulations).

In order to respect the restriction of a cold jet and test
particle shocks, we assume that the non-thermal particle energy
density is smaller than the thermal one, fixing the ratio between
them to 0.1 very close to the shock surface. Accounting for this,
and under the conditions of the shocks considered here, i.e., a
power similar to that of the jet going to thermal particles
downstream, the injected luminosity in the form of non-thermal
particles is $\sim 0.1\times L_{\rm j}$.

\subsubsection{Radiation}\label{rad}

{\bf Leptonic emission}

We can compute the SED of the synchrotron and IC components adopting a homogeneous model for the emitting
region. Since the magnetic field cannot be determined, two $B$-values, rendering different but
representative situations, have been adopted: $B=0.4$ and 50~G. The first value yields a dominant very high
energy IC component reaching energies well above 1~TeV. The second value corresponds to a high magnetic
field being in the limits of the hydrodynamical
approximation. This latter case implies dominant synchrotron radiation. 
To compute the impact of synchrotron self-absorption, $D_{\rm j,S}$ has been adopted as the 
width of the emitting region.

For $B=50$~G, $U_*\approx 300$~erg~cm$^{-3}$, and $\eta\sim 2\times 10^{-2}$, synchrotron energy losses
limit $E_{\rm max}\sim 1$~TeV, and the energy in non-thermal particles are radiated with similar
luminosities via synchrotron and IC channels due to comparable timescales. Above the injection energy, the
(cooled) particle spectrum follows a power-law of exponent $-$2 up to the energy at which convection and
radiative timescales become equal, around 1~GeV. 
In the one-zone approximation, electrons below 1~GeV escape the emitter before cooling
completely, although they could still produce radiation outside the binary system.
Above this energy, the particle spectrum becomes radiatively cooled
with index $-$3. 

\begin{figure}[]
\resizebox{\hsize}{!}{\includegraphics{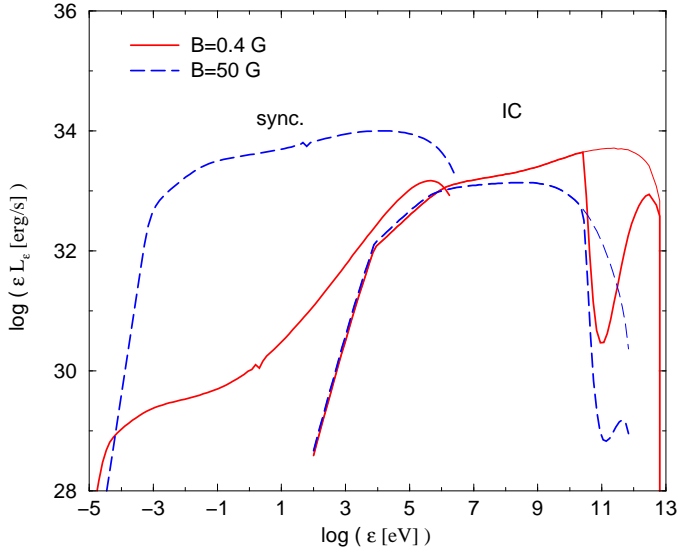}} \caption{Typical
computed SED of the synchrotron and IC emission produced in a
shock region. We adopt $B=0.4$ (solid line) and 50~G (long-dashed
line). The production (thin line) and the photon-photon absorbed (thick line)
SED cases are presented.} \label{sed}
\end{figure}

We have also explored the case of $B=0.4$~G, for which the maximum energy reaches around 10~TeV. In this
case, low magnetic fields render a synchrotron component with luminosities well below the IC ones and
a hard particle spectrum at the highest energies due to the impact of KN IC losses, which  harden the
particle energy distribution when dominant (on the contrary, synchrotron and Thomson IC losses  steepen the
spectrum; see Khangulyan et al. \cite{khangulyan07} for a thorough discussion on this). We show in
Fig.~\ref{sed} the corresponding SED for these two cases. The total radiated luminosity is $\sim
10^{35}$~erg~s$^{-1}$, being shared by the synchrotron and the IC components with their proportion depending
on $B$. It is worth noting that the age of the source producing the SED presented in Fig.~\ref{sed} has been
fixed to the typical lifetime of the shocks ($\sim R_{\rm orb}/V_{\rm j}$).

We note that for $B=50$~G, when synchrotron emission is high, the synchrotron self-absorption frequency is
$\sim 60$~GHz when assuming a low energy cutoff at $\sim m_{\rm e}c^2$, effectively yielding little
radiation in the radio band. If the cutoff were at higher energies, electrons would radiate with a
monoenergetic particle distribution in the radio band for a such magnetic field, rendering a very hard
SED (of exponent 4/3) and little radiation in the radio band again. Therefore, it is unclear whether radio emission in
this scenario is detectable coming from the accelerator/emitter. Electron cooling and transport with the
flow bulk may yield extended synchrotron radiation in the radio band in case radio emitting electrons were
carried to farther regions of appropriate $B$.

IC scattering and photon-photon absorption in the stellar radiation field have been considered
using angular averaged cross sections. Since we do not apply this model to any specific object at present,
we do not account for the geometry of these interactions. Accounting for the geometry of photon-photon absorption, 
the latter can have less impact for certain accelerator/emitter located locations ($\ga 10^{12}$~cm) and 
orbital phases. This would improve the detection chance for this kind of sources in the TeV range.
We refer to Khangulyan et~al. (\cite{khangulyan07}) for an exhaustive discussion of IC and photon-photon
absorption angular effects. Secondary pairs produced in the system may also be relevant from the radiative
point of view (see Khangulyan et~al. \cite{khangulyan07}; Bosch-Ramon, Khangulyan \& Aharonian
\cite{bosch08}). 

{\bf Hadronic emission}

In case of protons, the densities in the jet are in general low and proton-proton energy losses\footnote{Other hadronic channels, like synchrotron proton,
photo-disintegration, or photo-meson production, have little efficiency or an extremely high threshold energy. They are not considered here.} are not very important,
yielding little amount of $\pi^0$-decay gamma-rays. The most suitable case is when proton escape velocity is low and density high in the region. This could occur in the
shell region. Recollimation shocks would lead to jet density enhancement and low convection velocities. Also, the mixing of jet and stellar wind material may also lead
to high density regions that could suffer the impact of relativistic protons.

In our context, gamma-ray fluxes $\ga 10^{32}$~erg~s$^{-1}$ would be expected from proton-proton interactions if $n_{\rm t}/V_{\rm d}\ga 10$ (i.e. $V_{\rm d}\sim
10^9$~cm~s$^{-1}$ implies $n_{\rm t}\ga 10^{10}$~cm$^{-3}$) for $L_{\rm p}=10^{35}$~erg~s$^{-1}$, where $L_{\rm p}$ and $n_{\rm t}$ are the relativistic proton
luminosity and the target density, respectively.  Although the efficiency of hadronic processes is relatively low for the production of gamma-rays, the interaction of
powerful jets and dense winds could indeed lead to the generation of detectable TeV emission. In these proton-proton interactions, neutrinos (e.g. Romero et al.
\cite{romero05}; Aharonian et~al. \cite{aharonian06a}; Christiansen, Orellana \& Romero \cite{christiansen06}) and secondary electron-positron pairs would also be
produced, with luminosities similar to the gamma-ray ones. The radiation of secondary pairs would be relevant mainly at low energies (e.g. Orellana et~al.
\cite{orellana07}). 

{\bf Variability}

Initially, the forward
and reverse shocks can produce non-thermal radiating particles as long as the jet goes through the system,
with timescales $\sim R_{\rm orb}/V_{\rm j}\approx 300$~s. In case recollimation shocks in the jet due to
cocoon material pressure formed, particle acceleration would last as long as the cocoon pressure were higher
than that of the jet. 
Since the cocoon pressure goes roughly like $\propto 1/t^2$ once the jet head
leaves the binary system, cocoon recollimation shocks inside the system should not last longer than $\sim R_{\rm
orb}/V_{\rm j}$. All these events would appear as short bursts of X-ray and even GeV--TeV emission, and
perhaps also in the radio band. 

Irregular jet ejection on timescales $> R_{\rm orb}/V_{\rm w}$ (the wind replenishment timescale) in a massive system could lead to (recurrent?) very high energy flares
as that observed in Cygnus~X-1 (Albert et~al. \cite{albert07}) or apparently in certain orbital phases in LS~5039 (Aharonian et~al. \cite{aharonian06b}) and
LS~I~+61~303 (Rico \cite{rico07}). The observed $\sim$~hour scale variability associated to these flares could be explained by a new ejection
facing stellar wind replenished during a previous jet inactivity period of similar or longer duration. Eventual recollimation shocks due to wind ram pressure instead of cocoon
pressure could stand as long as jet injection lasted and wind properties were relatively smooth. The variability of emission in this phase would be likely dominated by
jet injection fluctuations or wind inhomogeneities, with associated timescales $\ga R_{\rm j}/c$.

\subsection{The fate of the jet}

As a test for jet bending, numerical simulations in slab geometry have been performed. It has been shown that the wind can efficiently transport shell and cocoon
material sidewards. In the same way, the wind can significantly deflect the supersonic jet itself via transferring of enough lateral momentum. We have shown that even
in the case of a jet with kinetic power $10^{36}\,\rm{erg\,s^{-1}}$, the wind seems able to disrupt the flow. Such a disrupted jet may look like a radio structure
changing its morphology along the orbit, like that found by Dhawan et al. \cite{dhawan06} in LS~I~+61~303 and interpreted by these authors as the collision between the
star and a pulsar winds. This interpretation would be challenged by the morphologies obtained from hydrodynamical simulations by Romero et~al. (\cite{romero07b}) and
Bogovalov et al. (\cite{bogovalov07}).

The simulations and calculations presented here show that the recollimation shocks can, on one hand, easily develop inside the binary system region and, on the other
hand, may influence significantly the evolution of the jet. As already stated, these shocks may generate efficient radiation and also could favor jet mass loading and
deceleration via pinching, thus making deflection by the wind easier. These shocks were already reported in a work by Peter \& Eichler (\cite{pei95}), in which the
collimation of jets by the inertia of the ambient medium was studied. In the maps shown in that work, shocks similar to that observed in Fig.~\ref{fig:pow} are
observed. 

The jet parameters adopted here result in very small growth rates of Kelvin-Helmholtz instability, mainly in
the cases II and III. Therefore, disruption due to the growth of such an instability inside the region of
interaction with the wind is not expected. Otherwise, our simulations show that the jets develop non-linear
structures due to their initial overpressure, thus the growth of linear instabilities is not expected either. In
short, the stability of the jet depends mainly on its inertia as compared to that of the wind and on the
strength of the recollimation shocks (Perucho et al. \cite{pe+05,pm07}).

\section{Summary}\label{sum}

The crossing of a hydrodynamical supersonic jet through the inner
parts of an HMMQ renders strong "forward/reverse" and
recollimation shocks. These shocks produce strong pressure and
density enhancements, and some jet deceleration.  The impact of
the wind ram pressure on weak jets can prevent the formation of
larger scale jets because of strong bending and jet disruption. We
estimate that this occurs for jet kinetic luminosities below
$10^{36}$~erg~s$^{-1}$ for the strong winds of O type stars.

If efficient leptonic and/or hadronic particle acceleration takes place, the
presence of photon, matter and magnetic fields, will yield significant amounts of
radiation from radio to very high energies.
On one hand, leptons would radiate mainly via synchrotron and IC
processes. On the other hand, protons could interact with shell
nuclei producing gamma-rays via neutral pion decay, and
electron-positron pairs and neutrinos via decay of charged pions.
Secondary pairs may also yield some amount of low energy
radiation. In addition, the complexity of the wind/jet
interactions leads to a plethora of variability timescales of the
emission that could be more relevant than any intrinsic jet
fluctuation.

Further three-dimensional numerical simulations including more
realistic implementations of the problem, magnetic fields, and
radiative processes, should improve our understanding of the
physics of jets in microquasars.

\begin{acknowledgements}
The authors benefited from valuable discussions with D.
Khangulyan. The authors thank G.~E. Romero for a thorough reading of the manuscript. 
V.B-R. acknowledges support
by DGI of MEC under grant AYA2007-68034-C03-01 and FEDER funds.
V.B-R. gratefully acknowledges support from the Alexander von
Humboldt Foundation. This work was supported in part by the
Spanish \emph{Direcci\'on General de Ense\~nanza Superior} under
grant AYA2004-08067-C03-01. M.P. acknowledges support from a
postdoctoral fellowship of the \emph{Generalitat Valenciana}
(\emph{Beca Postdoctoral d'Excel$\cdot$l\`encia}).
\end{acknowledgements}

{}


\begin{thebibliography}{}

\bibitem[1996]{aharonian96}
Aharonian, F.~A. \& Atoyan, A.~M. 1996, SSRv, 75, 357

\bibitem[1998]{aharonian98}
Aharonian, F.~A. \& Atoyan, A.~M. 1996, NewAR, 42, 579

\bibitem[2006a]{aharonian06a}
Aharonian, F.~A., Anchordoqui, L.~A., Khangulyan, D., \& Montaruli,
T. 2006a, J. Phys. Conf. Ser., 39, 408

\bibitem[2006b]{aharonian06b}
Aharonian, F., Akhperjanian, A.~G., \& Bazer-Bachi, A.~R., et~al.
2006b, A\&A, 460, 743

\bibitem[2007]{albert07}
Albert, J., Aliu, E., Anderhub, H., et~al. 2007, ApJL, 665,51

\bibitem[1999]{alo99}
Aloy, M.~A., Ib\'a\~nez, J.M., Mart\'{\i}, J.~M., G\'omez, J.L., \&
M\"uller, E. 1999, ApJL, 523, 125

\bibitem[2007]{araudo07}
Araudo, A.~T., Romero, G.~E., Bosch-Ramon, V., \& Paredes, J.~M.
2007, A\&A, 476, 1289

\bibitem[1984]{bbr84}
Begelman, M.~C., Blandford, M.~D., \& Rees, M.~J. 1984, RvMP, 56, 225

\bibitem[1989]{bc89} Begelman, M.C. \& Cioffi, D.F. 1989, ApJL, 345,
21

\bibitem[1978]{bell78}
Bell, A. R. 1978 MNRAS, 182, 443

\bibitem[2007]{bogovalov07}
Bogovalov, S.~V., Khangulyan, D., Koldoba, A.~V., Ustyugova, G.~V., \&
Aharonian F.~A. 2007, MNRAS, submitted [astro-ph/0710.1961]

\bibitem[2005]{bosch05}
Bosch-Ramon, V., Aharonian, F. A., \& Paredes, J.~M. 2005, A\&A,
432, 609

\bibitem[2008]{bosch08}
Bosch-Ramon, V., khnagulyan, D., \& Aharonian, F.~A. 2008, A\&A, in press [astro-ph/0801.4547]

\bibitem[1992]{bykov92}
Bykov, A. M. \& Fleishman, G. D. 1992, MNRAS, 255, 269

\bibitem[2005]{cas05} 
Casares, J., Rib\'o, M., Ribas, I., Paredes, J.~M., Mart\'{\i}, J., \& Herrero,
A. 2005, MNRAS, 364, 899

\bibitem[2002]{corbel02}
Corbel, S., Fender, R.~P., \& Tzioumis, A.~K., et~al. 2002,
Science, 298, 196


\bibitem[2005]{corbel05}
Corbel, S., Kaaret, P., \& Fender, R.~P. 2005, et~al. ApJ, 632,
504

\bibitem[2006]{christiansen06}
Christiansen, H.~R., Orellana, M., \& Romero, G.~E. 2006, PhRvD, 73,
3012

\bibitem[2003]{derishev03}
Derishev, E.~V., Aharonian, F.~A., Kocharovsky, V.~V., \&
Kocharovsky, Vl.~V. 2003, PhRvD, 68, 3003

\bibitem[2006]{dhawan06}
Dhawan, V., Mioduszewski, A., \& Rupen, M. 2006, in Proc. of the
VI Microquasar Workshop, Como-2006

\bibitem[1983]{drury83}
Drury, L. 1983, RPPh, 46, 973

\bibitem[1949]{fermi49}
Fermi, E. 1949, PhysRev 75, 1169

\bibitem[2004]{gabici04}
Gabici, S. \& Blasi, P. 2004, APh, 20, 579

\bibitem[2005]{gallo05}
Gallo, E., Fender, R., \& Kaiser, C., et~al. 2005, Nature, 436, 819

\bibitem[1986]{gb86} Gies, D.~R. \& Bolton, C.~T. 1986, ApJ, 304,
371

\bibitem[2001]{gomez01}
G\'omez, J.~L. 2001, Ap\&SS, 276, 281

\bibitem[2003]{hardee03}
Hardee, P.~E. \& Hughes, P.~A. 2003, ApJ, 583, 116

\bibitem[2003]{heindl03}
Heindl, W.~A., Tomsick, J.~A., Wijnands,~R., \& Smith, D.~M. 2003, ApJ, 588, L97 

\bibitem[2002]{heinz02}
Heinz, S. 2002, A\&A, 388, L40

\bibitem[2006]{heinz06}
Heinz, S. 2006, ApJ, 636, 316

\bibitem[2002]{heinzS02}
Heinz, S. \& Sunyaev, R. 2002, A\&A, 390, 751

\bibitem[1996]{kang96}
Kang, H., Ryu, D., \& Jones, T.~W. 1996, ApJ, 56, 422

\bibitem[2008]{khangulyan07}
Khangulyan, D., Aharonian, F.~A., \& Bosch-Ramon, V. 2008, MNRAS, 383, 467

\bibitem[1998]{kofa98}
Komissarov, S.~S. \& Falle, S.~A.~E.~G. 1998, MNRAS, 297, 1087

\bibitem[2005]{kra05}
Krause, M. 2005, A\&A, 431, 45

\bibitem[1962]{laan62}
van der Laan, H. 1962, MNRAS, 124, 179

\bibitem[1991]{lea91}
Leahy, J.~P. 1991, in \emph{Beams and Jets in Astrophysics}, ed.
P.A. Hughes, Cambridge Astrophysics Series, p.100

\bibitem[2005]{leis05}
Leismann, T., Ant\'on, L., Aloy, M.~A., M\"uller, E., Mart\'{\i},
J.~M., Miralles, J.~A., \& Ib\'a\~nez, J.~M. 2005, A\&A, 436, 503

\bibitem[2002]{marshall02}
Marshall, H.~L., Canizares, C.~R., \& Schulz, N.~S. 2002, ApJ, 564,
941

\bibitem[1996]{marti96}
Mart\'i, J., Rodriguez, L.~F., Mirabel, I.~F., \& Paredes, J.~M. 1996, A\&A, 306, 449

\bibitem[1997]{mart97}
Mart\'{\i}, J.M., M\"uller, E., Font, J.A.,
Ib\'a\~nez, J.M., \& and Marquina, A. 1997, ApJ,
479, 151

\bibitem[1999]{mirabel99}
Mirabel, I.~F. \& Rodr\'iguez, L.~F. 1999, AR\&A, 37, 409

\bibitem[2007]{mizu07}
Mizuno, Y., Hardee, P.E., \&  Nishikawa, K.-I. 2007, ApJ, 662, 835

\bibitem[2007]{orellana07}
Orellana, M., Bordas, P., Bosch-Ramon, V., Romero, \& G.~E., Paredes,
J.~M. 2007, A\&A, 476, 9

\bibitem[2006]{owocki06}
Owocki, S. \& Cohen, D. 2006, ApJ, 648, 565

\bibitem[2006]{paredes06}
Paredes, J.~M., Bosch-Ramon, V., \& Romero, E. 2006, A\&A, 451, 259

\bibitem[2007]{paredes07}
Paredes, J.~M., Rib\'o, M., \& Bosch-Ramon, V., et~al. 2007, ApJL, 664, 39

\bibitem[2005]{pe+05}
Perucho, M., Mart\'{\i}, J.~M., \& Hanasz, M. 2005, A\&A, 443, 863

\bibitem[2006]{pe+06}
Perucho, M., Lobanov, A.~P., Mart\'{\i}, J.~M., \& Hardee, P.~E. 2006,
A\&A, 456, 493

\bibitem[2007]{pm07}
Perucho, M. \& Mart\'{\i}, J.~M. 2007, MNRAS, in press,
arXiv:0709.1784

\bibitem[1995]{pei95}
Peter, W. \& Eichler, D. 1995, ApJ, 438, 244

\bibitem[1999]{protheroe99}
Protheroe, R. J. 1999, ADP-AT-98-9 [astro-ph/9812055]

\bibitem[2006]{puls06}
Puls, J., Markova, N., \& Scuderi, S., et~al. 2006, A\&A, 454, 625

\bibitem[1974]{rees74}
Rees, M.~J. \& Gunn, J.~E. 1974, MNRAS, 167, 1

\bibitem[2005]{ribo05}
Rib\'o, M. 2005, ASPC, 340, 269

\bibitem[2007]{rico07}
Rico, J. for the MAGIC collaboration, talk presented in the conference:
High energy processes in relativistic jets, Dublin, Ireland

\bibitem[2004]{rieger04}
Rieger, F.~M. \& Duffy, P. 2004, ApJ 617, 155

\bibitem[2003]{romero03}
Romero, G.~E., Torres, D.~F., Kaufman Bernad\'o, M. M., \&
Mirabel, I.~F. 2003, A\&A, 410, L1,

\bibitem[2005]{romero05}
Romero, G.~E., \& Orellana, M. 2005, A\&A, 439, 237

\bibitem[2007a]{clumps}
Romero, G.~E., Owocki S.~P., Araudo, A.~T., Townsend, R., \& Benaglia
P. 2007a, workshop proceedings "Clumping in Hot Star Winds"
[astro-ph/0708.1525]

\bibitem[2007b]{romero07b}
Romero, G.~E., Okazaki, A.~T., Orellana, M. \& Owocki, S.~P.
2007b, A\&A, 474, 15 

\bibitem[1999]{ro99}
Rosen, A., Hughes, P.A., Duncan, G.C., \& Hardee, P.E. 1999, ApJ,
516, 729

\bibitem[2002]{sch02}
Scheck, L., Aloy, M.A., Mart\'{\i}, J.M.,
G\'omez, J.L., \& M\"uller, E. 2002, MNRAS, 331, 615

\bibitem[2006]{tudose06}
Tudose, V., Fender, R.~P., \& Kaiser, C.~R., et al. 2006, MNRAS,
372, 417

\bibitem[2000]{velazquez00}
Vel\'azquez, P.~F. \& Raga, A.~C. 2000, A\&A, 362, 780

\bibitem[1980]{zealey80}
Zealey, W. J., Dopita, M. A., \& Malin, D. F. 1980, MNRAS, 192,
731
\end{thebibliography}
\end{document}